\documentclass[12pt,a4paper]{article}

\usepackage[cp1251]{inputenc}
\usepackage[T2A]{fontenc}
\usepackage[russian]{babel}

\usepackage[dvips]{graphicx}
\usepackage{amsmath}
\usepackage{amssymb}
\usepackage{amsxtra}
\usepackage{amsthm}
\usepackage{latexsym}
\usepackage{array}
\usepackage{multirow}
\usepackage{hhline}
\usepackage[in]{fullpage}
\usepackage{indentfirst}

\newcommand{\rank}{\mathop{\rm rank}\nolimits}

\usepackage{hyperref}
\theoremstyle{plain}
\newtheorem{theorem}{Теорема}

\begin{document}
\begin{flushleft}
УДК 532.5.031, 517.938.5
\end{flushleft}
\begin{center}

О БИФУРКАЦИИ ЧЕТЫРЕХ ТОРОВ ЛИУВИЛЛЯ В ОДНОЙ ОБОБЩЕННОЙ ИНТЕГРИРУЕМОЙ МОДЕЛИ ВИХРЕВОЙ ДИНАМИКИ

П.~Е. Рябов$^{1,2,3}$

${}^{1}$\! Финансовый университет при Правительстве Российской Федерации \\
125993, Россия, г.~Москва, Ленинградский проспект, д.~49

${}^{2}$\! Институт машиноведения РАН~им.~А.~А.~Благонравова \\
119334, Россия, г.~Москва, ул.~Бардина, д.~4

${}^{3}$\! Удмуртский государственный университет \\
426034, Россия, г.~Ижевск, ул.~Университетская, д.~1

E-mail: PERyabov@fa.ru

\end{center}



\section{Введение}

Обобщенная математическая модель динамики двух точечных вихрей в бо\-зе-эйн\-штей\-нов\-ском кон\-ден\-са\-те, заключенном в гармонической ловушке, и  динамики двух точечных вихрей в идеальной жидкости, ограниченной круговой областью, описывается системой дифференциальных уравнений, которая может быть представлена в гамильтоновой форме
\begin{equation}
\label{x1}
\displaystyle{\Gamma_k\dot x_k=\frac{\partial H}{\partial y_k} (z_1,z_2);\quad \Gamma_k\dot y_k=-\frac{\partial H}{\partial x_k} (z_1,z_2),\quad k=1,2,}
\end{equation}
где гамильтониан $H$ имеет вид:
\begin{equation}
\label{x2}
\begin{array}{l}
\displaystyle{H=\frac{1}{2}\Bigl[\Gamma_1^2\ln(1-|z_1|^2)+\Gamma_2^2\ln(1-|z_2|^2)+}\\[5mm]
\displaystyle{+\Gamma_1\Gamma_2\ln\left(\frac{[|z_1-z_2|^2+(1-|z_1|^2)(1-|z_2|^2)]^\varepsilon}{|z_1-z_2|^{2(c+\varepsilon)}}\right)\Bigr].}
\end{array}
\end{equation}
Здесь через $z_k=x_k+{\rm i}y_k$ обозначены декартовы координаты $k$-ого вихря ($k=1,2$) с интенсивностями $\Gamma_k$. Физический параметр $``c``$ выражает собой меру вихревого взаимодействия, $\varepsilon$ -- параметр деформации, который характеризует два предельных случая, а именно, при $\varepsilon=0$ -- модель двух точечных вихрей в бо\-зе-эйн\-штей\-нов\-ском кон\-ден\-са\-те, заключенном в гармонической ловушке \cite{kevrekPhysLett2011}, \cite{kevrek2013}, \cite{kevrikidis2014}, а при $c=0$, $\varepsilon=1$ -- модель двух точечных вихрей в идеальной жидкости, ограниченной круговой областью \cite{Greenhill1877}, \cite{BorMamSokolovskii2003}, \cite{BorMam2005book}.

Фазовое пространство $\cal P$ задается в виде прямого произведения двух открытых кругов радиуса $1$ с выколотом множеством столкновениий вихрей:
\begin{equation*}
{\cal P}=\{(z_1,z_2)\,:\, |z_1|<1,\,|z_2|<1,z_1\ne z_2\}.
\end{equation*}
Пуассонова  структура на фазовом пространстве $\cal P$ задается в стандартном виде
\begin{equation}
\label{x4}
\{z_k,\bar{z}_j\}=-\frac{2\rm i}{\Gamma_k}\delta_{kj},
\end{equation}
где $\delta_{kj}$ -- символ Кронекера.

Система $\eqref{x1}$ допускает один дополнительный первый интеграл движения -- \textit{момент завихренности}:
\begin{equation*}
F=\Gamma_1|z_1|^2+\Gamma_2|z_2|^2.
\end{equation*}

Функция $F$ вместе с гамильтонианом $H$ образуют на $\cal P$ полный инво\-лю\-тив\-ный набор интегралов системы $\eqref{x1}$.
Согласно теореме Лиувилля-Арнольда  регулярная поверхность уровня первых интегралов вполне интегрируемой гамильтоновой системы представляет собой несвязное объединение торов, заполненных условно-периодическими траекториями. Определим \textit{интегральное отображение} ${\cal F}\,:\, {\cal P}\to {\mathbb R}^2$, полагая $(f,h)={\cal F}(\boldsymbol x)=(F(\boldsymbol x), H(\boldsymbol x))$.  Обозначим через $\cal C$ совокупность всех критических точек отображений момента, то есть точек, в которых $\rank d{\cal F}(\boldsymbol x) < 2$. Множество критических значений $\Sigma = {\cal F}({\cal C}\cap{\cal P})$ называется \textit{бифуркационной диаграммой}.

В работах \cite{SokRyabRCD2017} и \cite{sokryab2018} при определенном значении параметра вихревого взаимодействия ($c=1$)  в случае интенсивностей противоположных и одинаковых знаков  аналитически исследована бифуркационная диаграмма задачи о движении системы двух точечных вихрей в бозе-эйнштейновском конденсате. В \cite{RyabDan2019apper} и \cite{RyabSocND2019appear} выполнена редукция к системе с одной степенью свободы и при отсутствии параметра деформации ($\varepsilon=0$)  для  значениях физического параметра $c>3$ обнаружена бифуркация трёх торов в один. Такая бифуркация оказалась неустойчивой и приведено её возмущённое слоение. Для другого предельного случая ($c=0, \varepsilon=1$) бифуркационный анализ динамики двух точечных вихрей в идеальной жидкости, ограниченной круговой областью, выполнен в \cite{BorMamSokolovskii2003} и  \cite{BorMam2005book}.  Для указанных предельных случаев были получены совершенно различные бифуркационные диаграммы. А.\,В.~Борисов предложил рассмотреть обе эти интегрируемые модели и выяснить, как связаны бифуркационные диаграммы обоих предельных случаев. В настоящей публикации в случае положительной вихревой пары ($\Gamma_1=\Gamma_2=1$), представляющий интерес для физических экспериментальных приложений, для обобщенной математической модели, описываемой \eqref{x1}, \eqref{x2}, получена новая бифуркационная диаграмма, для которой указана бифуркация четырех торов в один. Наличие бифуркаций трех и четырех торов в интегрируемой модели динамики вихрей, имеющих положительные интенсивности, свидетельствует о сложном переходе и связи  бифуркационных диаграмм обоих предельных случаев.

\section{Бифуркационная диаграмма}
В случае положительной вихревой пары ($\Gamma_1=\Gamma_2=1$) определим полиномиальные выражения $F_k$ от фазовых переменных
\begin{equation*}
\begin{array}{l}
F_1=x_1y_2-y_1x_2,\\[3mm]
F_2=[x_1(x_2^2+y_2^2)-x_2]\Bigl\{(x_1^2+x_2^2)(x_2^2+y_2^2)[x_1(x_2^2+y_2^2)-cx_2]+\\[3mm]
x_2[(c-2)(x_2^2+y_2^2)^2x_1^2+cx_2^2]\Bigr\}+\\[3mm]
+\varepsilon(x_2^2+y_2^2-1)(x_1^2(x_2^2+y_2^2)-x_2^2)[(x_2^2+y_2^2)(x_1^2-x_1x_2+x_2^2)-x_2^2]
\end{array}
\end{equation*}
и обозначим через ${\cal N}_1$ и ${\cal N}_2$ замыкания множеств решений следующих систем
\begin{equation}
\label{y1}
x_1+x_2=0,\quad y_1+y_2=0
\end{equation}
и
\begin{equation}
\label{y11}
F_1 = 0,\quad F_2 = 0.
\end{equation}

Тогда справедлива теорема.

\begin{theorem}
\label{t1}
В случае положительной вихревой пары множество $\cal C$ критических точек  отображения момента $\cal F$ совпадает с множеством решений систем \eqref{y1} и \eqref{y11}. Множества ${\cal N}_1$ и ${\cal N}_2$   являются двумернымы инвариантными подмногообразиями  системы \eqref{x1} с гамильтонианом \eqref{x2}.
\end{theorem}

Для определения бифуркационной диаграммы $\Sigma$ удобно перейти к полярным координатам:
\begin{equation}
\label{y2}
x_1 = r_1\cos\theta_1,\quad y_1 = r_1\sin\theta_1,\quad
x_2 = r_2\cos\theta_2,\quad y_2 = r_2\sin\theta_2.
\end{equation}

Подстановка \eqref{y2} в \eqref{y1} и \eqref{y11} приводит к системе
\begin{equation*}
\left\{\begin{array}{l}
\theta_1=\theta_2+\pi;\\
\left[\begin{array}{l}
r_1=r_2;\\
(1+r_1r_2)[(r_1^2+r_2^2)(r_1r_2+c)-(c-2)r_1^2r_2^2-c]+\\[3mm]
+\varepsilon(1-r_1^2)(1-r_2^2)(r_1^2+r_1r_2+r_2^2-1)=0.
\end{array}\right.
\end{array}\right.
\end{equation*}

Бифуркационная диаграмма $\Sigma$ определена  на  плоскости
${\mathbb R}^2(f,h)$ и состоит из двух кривых $\gamma_1$ и $\gamma_2$, где
\begin{equation}\label{y4}
\begin{array}{l}
\displaystyle{\gamma_1: h=\ln\Bigl(1-\dfrac{f}{2}\Bigr)-\frac{1}{2}(c+\varepsilon)\ln(2f)+\varepsilon\ln\left(1+\frac{f}{2}\right),\quad 0<f<2;}\\[3mm]
\gamma_2: \left\{
\begin{array}{l}
\displaystyle{h=\ln\Bigl(x^\varepsilon\sqrt{x^2-1}z^{1-c}\Bigr),}\\[3mm]
\displaystyle{f=z^2-2xz+2,} \\[3mm]
\displaystyle{z=\frac{x[(\varepsilon+c)(x^2-1)+1]}{(\varepsilon+1)x^2-\varepsilon}},
\end{array}\right. x\in (1; x_0].
\end{array}
\end{equation}
Здесь через $x_0$ обозначен корень уравнения
\begin{equation}\label{y5}
  (z-2x)^2=4(x^2-1),\quad  x>1.
\end{equation}

На рис.~1 и рис.~2 в случае положительной вихревой пары для значений параметров $\varepsilon =28, c=12$ приведена бифуркационная диаграмма и ее увеличенный фрагмент. Отметим, что кривая $\gamma_2$ имеет точки возврата $A, B$ и точку касания $C$ с кривой $\gamma_1$ для указанных значений параметров при
\begin{equation*}
\begin{array}{l}
\displaystyle{x_0=\frac{\sqrt{570}}{1140}\sqrt{2312+\sqrt[3]{2885048-294690\sqrt{6}}+\sqrt[3]{2885048+294690\sqrt{6}}}\approx1,06678;}\\[3mm]
\displaystyle{f_C=\frac{2}{57}\Bigl[45+\frac{\sqrt[3]{141^2}}{\sqrt[3]{45+19\sqrt{6}}}-\sqrt[3]{141(45+19\sqrt{6})}\Bigr]\approx 0,9667958154;}\\[3mm]
h_C\approx -2,8066772742.
\end{array}
\end{equation*}
Указанная  на рис.~2~a) точка касания $C$ удовлетворяет \eqref{y5}.

\begin{figure}[!ht]
\centering
\includegraphics[width=0.5\textwidth]{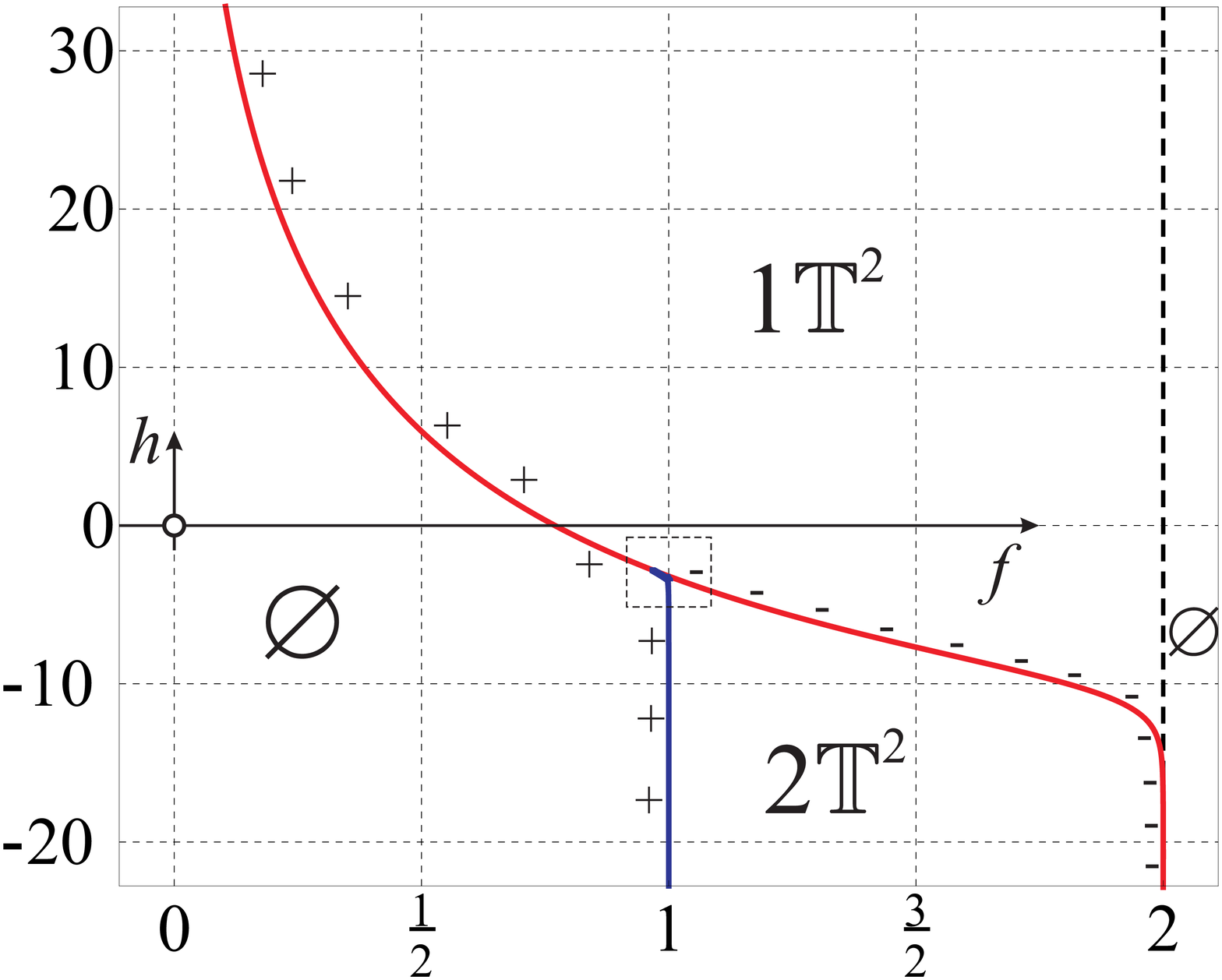}
\caption{Бифуркационная диаграмма $\Sigma$.}
\label{fig1}
\end{figure}
\begin{figure}[!ht]
\centering
\includegraphics[width=1\textwidth]{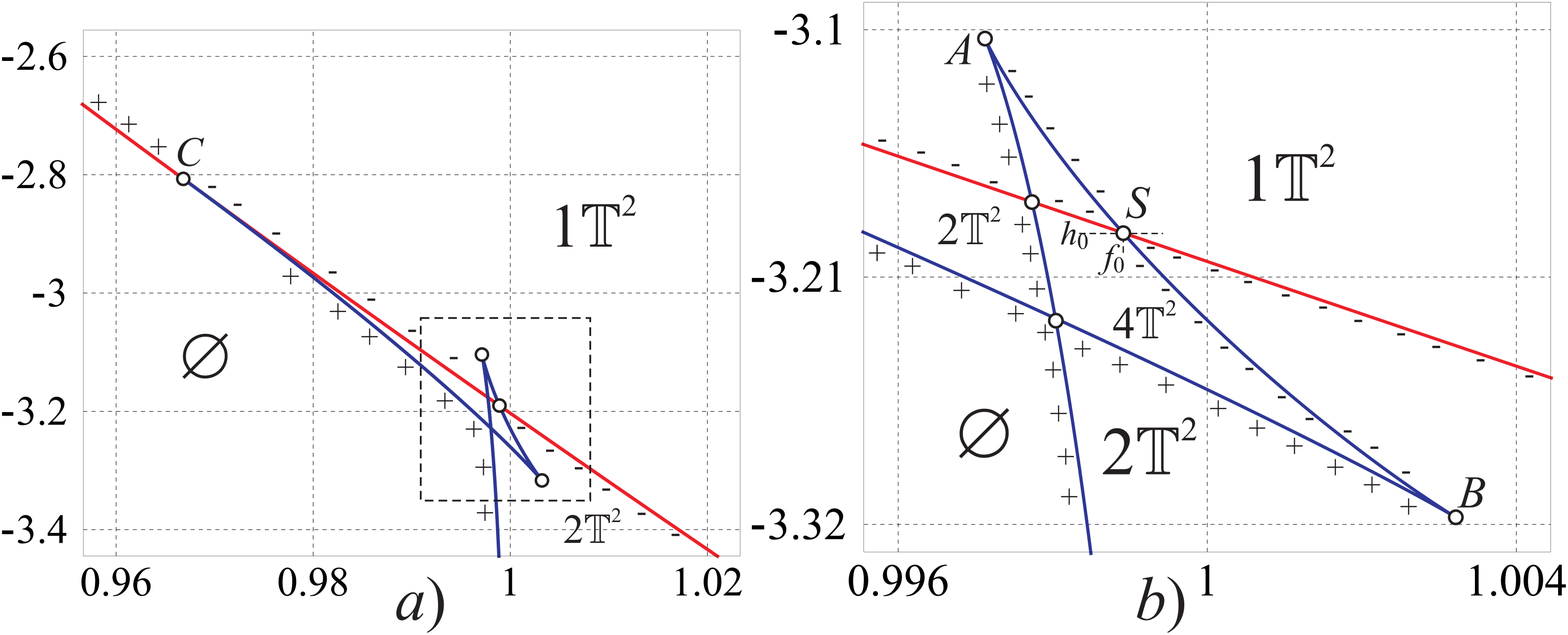}
\caption{Увеличенный фрагмент бифуркационной диаграммы.}
\label{fig2}
\end{figure}

Знаки $``+``$ и $``-``$ соответствуют эллиптическим (устойчивым) и гиперболическим периодическим решениям в фазовом пространстве \cite{BolBorMam1}. Как и следовало ожидать, смена типа происходит в точках возврата $A$ и $B$, а также в точке касания $C$ бифуркационной диаграммы $\Sigma$.
Для наглядности, приведем явное выражение коэффициента $C_F$, который отвечает за тип (эллип\-ти\-че\-ский/ги\-пер\-бо\-ли\-че\-ский) гладкой ветви кривой $\gamma_1$:
\begin{equation*}\label{y_6}
\begin{array}{l}
\gamma_1: C_F=(4-c+3\varepsilon)f^3+2(c+4-7\varepsilon)f^2+4(c+5\varepsilon)f-8(c+\varepsilon),\quad 0<f<2.
\end{array}
\end{equation*}
При $C_F < 0$ получим точку типа ``центр'' (соответствующее периодическое решение имеет эллиптический тип, является устойчивым периодическим решением в фазовом пространстве, пределом концентрического семейства двумерных регулярных торов), а при $C_F > 0$ получим точку типа ``седло'' (соответствующее периодическое решение имеет гиперболический тип, существуют движения, асимптотические к этому решению, лежащие на двумерных сепаратрисных поверхностях).

\section{О бифуркации  четырех торов}
Выполним явное приведение к системе с одной степенью свободы в случае положительной вихревой пары ($\Gamma_1=\Gamma_1=1$) подобно тому, как это сделано в \cite{RyabDan2019apper}. Для этого в системе \eqref{x1} с гамильтонианом  \eqref{x2} перейдем от фазовых переменных $(x_k,y_k)$ к новым переменным $(u,v,\alpha)$ по формулам:
\begin{equation*}\label{z1}
\begin{array}{l}
x_1=u\cos(\alpha)-v\sin(\alpha),\quad y_1=u\sin(\alpha)+v\cos(\alpha),\\[3mm]
\displaystyle{x_2=\sqrt{f-u^2-v^2}\cos(\alpha),\quad y_2=\sqrt{f-u^2-v^2}\sin(\alpha).}
\end{array}
\end{equation*}

Физические переменные $(u,v)$ представляют собой декартовы координаты одного из вихрей в системе координат, связанной с другим вихрем, вращающейся вокруг центра завихренности. Обратная замена
\begin{equation*}
U=\frac{x_1x_2+y_1y_2}{\sqrt{x_2^2+y_2^2}},\quad V=\frac{y_1x_2-x_1y_2}{\sqrt{x_2^2+y_2^2}}
\end{equation*}
приводит к каноническим переменным относительно скобки \eqref{x4}:
\begin{equation*}
\{U,V\}=-\{V,U\}=1,\quad \{U,U\}=\{V,V\}=0.
\end{equation*}

Система по  отношению к новым переменным $(u,v)$ является гамильтоновой
\begin{equation}\label{z2}
\dot u=\frac{\partial H_1}{\partial v},\quad
\dot v=-\frac{\partial H_1}{\partial u}
\end{equation}
с гамильтонианом
\begin{equation}
\label{z3}
\begin{array}{l}
\displaystyle{H_1=\frac{1}{2}\Bigl\{\ln[(1-u^2-v^2)(1-f+u^2+v^2)]-(c+\varepsilon)\ln(f-2u\sqrt{f-u^2-v^2})+}\\[3mm]
\displaystyle{+\varepsilon\ln[(1-u\sqrt{f-u^2-v^2})^2+v^2(f-u^2-v^2)]\Bigr\}}.
\end{array}
\end{equation}

Угол поворота $\alpha(t)$ вращающейся системы координат удовлетворяет дифференциальному уравнению
\begin{equation*}
\begin{array}{l}
\displaystyle{\dot\alpha=
\frac{1}{1-f+u^2+v^2}+c\frac{R_1(u,v)}{Q_1(u,v)}+\varepsilon\frac{R_2(u,v)}{Q_2(u,v)}},
\end{array}
\end{equation*}
где
\begin{equation*}
\begin{array}{l}
R_1(u,v)=f(u+\sqrt{f-u^2-v^2})-2u[v^2+u(u+\sqrt{f-u^2-v^2})],\\[3mm]
Q_1(u,v)=\sqrt{f-u^2-v^2}[(f-2u^2)^2+4u^2v^2],\\[3mm]
R_2(u,v)=(1-u^2-v^2)[\sqrt{f-u^2-v^2}(1+u^2+v^2)-u(1+f-u^2-v^2)],\\[3mm]
Q_2=\sqrt{f-u^2-v^2}\Bigl\{(u^2+v^2)f^2-2\sqrt{f-u^2-v^2}(1+u^2+v^2)(1+f-u^2-v^2)u+\\[3mm]
+f[1-(u^2+v^2)^2]+4u^2(f-u^2-v^2)\Bigr\}
\end{array}
\end{equation*}

Неподвижные точки редуцированной системы \eqref{z2} опре\-де\-ля\-ют\-ся кри\-ти\-че\-скими точками приведенного гамильтониана \eqref{z3} и соответствуют  относительным равновесиям вихрей в системе \eqref{x1}. Для фиксированного значения интеграла момента завихренности $f$ регулярные уровни приведенного гамильтониана -- компактны и движения происходят по замкнутым кривым.
Можно показать, что критические значения приведенного гамильтониана определяют бифуркационную диаграмму \eqref{y4}. В точке $S$  пересечения  бифуркационных кривых $\gamma_1$ и $\gamma_2$ (рис.~2 б)), для которой $x_S=1,008383; f_0=0,9989101; h_0=-3,1903429$, движение на плоскости $(u,v)$ происходит по кривой, которая топологически устроена как $\mathbb S^1\,\dot{\cup}\,\mathbb S^1\,\dot{\cup}\,\mathbb S^1\,\dot{\cup}\,\mathbb S^1$ (рис.~3~а), а интегральная критическая поверхность представляет собой тривиальое расслоение над $\mathbb S^1$ со слоем $\mathbb S^1\,\dot{\cup}\,\mathbb S^1\,\dot{\cup}\,\mathbb S^1\,\dot{\cup}\,\mathbb S^1$.

\begin{figure}[!ht]
\centering
\includegraphics[width=1\textwidth]{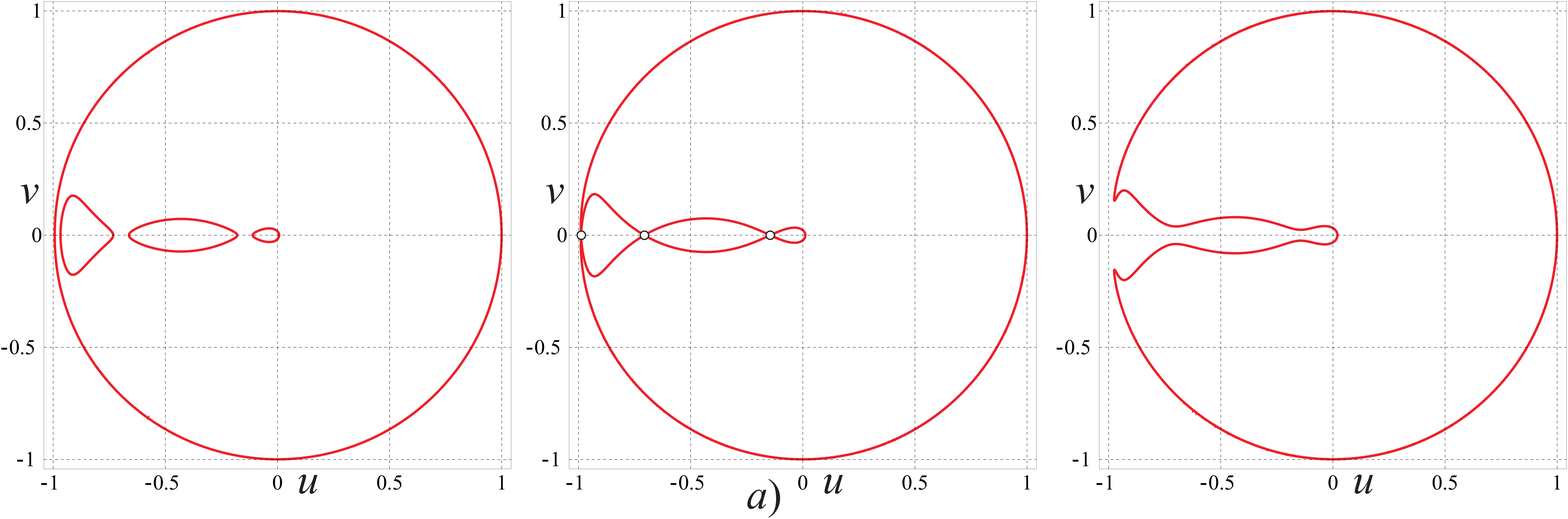}
\caption{Линии уровня приведенного гамильтониана $H_1$ вдоль прямой $h=h_0$.}
\label{fig4}
\end{figure}

При переходе через точку  $S$ бифуркационной диаграммы $\Sigma$ вдоль прямой $h=h_0$  (рис.~2 б) реализуется бифурка\-ция че\-ты\-рех то\-ров в один  $4\mathbb T^2 \to \mathbb S^1\times\left(\mathbb S^1\,\dot{\cup}\,\mathbb S^1\,\dot{\cup}\,\mathbb S^1\,\dot{\cup}\,\mathbb S^1\right)$ $\to \mathbb T^2$. С помощью линий уровней приведенного гамильтониана \eqref{z3} на рис.~3 наглядно продемонстрирована указанная бифуркация.

В заключении отметим, что аналитические результаты настоящей публикации (бифуркационная диаграмма \eqref{y4}, сведение к системе с одной степенью свободы \eqref{z2}, анализ устойчивости) в случае положительной вихревой пары ($\Gamma_1=\Gamma_2=1$) составляют основу компьютерного моделирования абсолютной динамики вихрей в неподвижной системе координат, описываемой \eqref{x1} и \eqref{x2}, в случае \textit{произвольных} значений интенсивностей $\Gamma_1, \Gamma_2$, физического параметра $c$  и параметра деформации $\varepsilon$.

Автор выражает благодарность А.\,В.\,Бо\-ри\-со\-ву за постановку задачи. Работа выполнена при поддержке гранта РФФИ № 17-01-00846.


\end{document}